\documentclass[sigplan,screen]{acmart}

\AtBeginDocument{\providecommand\BibTeX{{\normalfont B\kern-0.5em{\scshape i\kern-0.25em b}\kern-0.8em\TeX}}}

\usepackage{amsmath}
\usepackage{mathtools}
\usepackage{multirow}
\usepackage{pdflscape}
\usepackage{afterpage}
\usepackage{microtype}

\usepackage{theorems}[2015/07/09]

\providecommand*{\parensmathoper}[3][]{\ensuremath{\mathoper{#2}\ifempty{#3}{}{#1(#3#1)}}}

\providecommand*{\BBrackets}[2][]{\ifempty{#2}{}{\llbracket#2\rrbracket_{#1}}}

\providecommand*{\true}{\mathit{true}}
\providecommand*{\false}{\mathit{false}}

\renewcommand*{\implies}{\Rightarrow}
\providecommand*{\eg}   {e.g.}
\providecommand*{\ie}   {i.e.,} 
\providecommand*{\resp} {respectively}

\makeatletter
\providecommand{\ifempty}[3]{\def\@@@temp{#1}\ifx\@@@temp\@empty#2\else#3\fi}
\makeatother

\newcommand*{\Psyms}{\Pi}
\newcommand{\fret}{\textsc{Fret}}
\newcommand{\Fret}{\fret}

\newcommand{\fretish}{{FRETish}}

\newcommand{\mtl}{MTL}
\newcommand{\MTL}{\mtl}
\newcommand{\PVS}{PVS}
\newcommand*{\N}{\ensuremath{\mathbb{N}}}

\newcommand{\B}{\mathbb{B}}

\newcommand{\previous}{\mathoper{\mathcal{Y}}}
\newcommand{\historically}[2][I]{\mathoper{\mathcal{H}_{#1}}\ifempty{#2}{}{#2}}
\newcommand{\once}[2][I]{\mathoper{\mathcal{O}_{#1}}\ifempty{#2}{}{#2}}
\newcommand{\since}[3][I]{\ifempty{#2}{\mathcal{S}_{#1}}{#2 \mathop{\mathcal{S}_{#1}} #3}}
\newcommand{\sir}[2]{\ifempty{#1}{\mathcal{S}_{\mathit{inc}}^{\mathit{req}}}
{#1 \mathop{\mathcal{S}_{inc}^{req}} #2}}
\newcommand{\sio}[2]{\ifempty{#1}{\mathcal{S}_{\mathit{inc}}^{\mathit{opt}}}
{#1 \mathop{\mathcal{S}_{inc}^{opt}} #2}}
\newcommand{\ra}{\rightarrow}

\newcommand*{\TraceDom}{\ensuremath{\mathbb{T}}}
\newcommand{\MTLDom}{\mathbb{MTL}}
\newcommand{\fieldscope}{{\textit{scope}}}
\newcommand{\fieldcondition}{{\textit{condition}}}
\newcommand{\fieldtiming}{{\textit{timing}}}
\newcommand{\fieldresponse}{{\textit{response}}}
\newcommand{\fieldcomponent}{{\textit{component}}}
\newcommand{\fieldshall}{{\textit{shall}}}
\newcommand{\fieldsatisfy}{{\textit{satisfy}}}

\newcommand{\fretafter}{\parensmathoper{\mathit{after}}}
\newcommand{\afterscope}{\fretafter}
\newcommand{\fretbefore}{\parensmathoper{\mathit{before}}}
\newcommand{\beforescope}{\fretbefore}
\newcommand{\fretin}{\parensmathoper{\mathit{in}}}
\newcommand{\inscope}{\fretin}
\newcommand{\fretnotin}{\parensmathoper{\mathit{notIn}}}
\newcommand{\notinscope}{\fretnotin}
\newcommand{\fretonlyafter}{\parensmathoper{\mathit{onlyAfter}}}
\newcommand{\onlyafterscope}{\fretonlyafter}
\newcommand{\fretonlybefore}{\parensmathoper{\mathit{onlyBefore}}}
\newcommand{\onlybeforescope}{\fretonlybefore}
\newcommand{\fretonlyin}{\parensmathoper{\mathit{onlyIn}}}
\newcommand{\onlyinscope}{\fretonlyin}
\newcommand{\fretnull}{\mathit{null}}
\newcommand{\nullcond}{\mathit{null}}
\newcommand{\nullscope}{\fretnull}
\newcommand{\fretonly}{{\textit{only}}}
\newcommand{\fretimmediately}{\mathit{immediately}}
\newcommand{\fretnext}{\mathit{next}}
\newcommand{\fretnever}{\mathit{never}}
\newcommand{\freteventually}{\mathit{eventually}}
\newcommand{\fretalways}{\mathit{always}}
\newcommand{\fretwithin}{\parensmathoper{\mathit{within}}}
\newcommand{\fretfor}{\parensmathoper{\mathit{for}}}
\newcommand{\fretuntil}{\parensmathoper{\mathit{until}}}
\newcommand{\fretaftertime}{\parensmathoper{\mathit{after}}}
\newcommand{\fretReq}[4]{\langle #1, #2, #3, #4 \rangle}

\newcommand{\oli}{\mathit{OLI}}
\newcommand{\LNI}[1]{{\dot{\mathbb{I}}}\ifempty{#1}{}{_{#1}}}
\newcommand{\lb}{\parensmathoper{\mathit{lb}}}
\newcommand{\ub}{\parensmathoper{\mathit{ub}}}
\newcommand{\neglni}[1]{\overline{#1}}
\newcommand{\mode}{\mathit{mode}}
\newcommand{\stopcond}{\mathit{stop}}
\newcommand{\cond}{\mathit{cond}}
\newcommand{\res}{\mathit{res}}
\newcommand{\modeInt}{\mathit{modeInt}}
\newcommand{\stopInt}{\mathit{stopInt}}
\newcommand{\condInt}{\mathit{condInt}}
\newcommand{\resInt}{\mathit{resInt}}

\newcommand{\emptylin}{\epsilon}

\newcommand{\boolSem}[2]{\mathcal{I}\ifempty{#1}{}{(#1,#2)}}
\newcommand{\scopeSem}[2][\rho]{\mathcal{S}\llbracket#2\rrbracket_{#1}}
\newcommand{\timeSem}[4][\rho]{\mathcal{T}\llbracket#2\rrbracket_{#1}(#3,#4)}
\newcommand{\triggers}[3]{\mathit{triggers}\ifempty{#1}{}{(#1,#2,#3)}}
\newcommand{\stops}[3]{\mathit{stops}\ifempty{#1}{}{(#1,#2,#3)}}
\newcommand{\firststop}[4]{\mathit{firststop}\ifempty{#1}{}{(#1,#2,#3,#4)}}

\newcommand{\scopeleft}{\parensmathoper{\mathit{left}}}
\newcommand{\scoperight}{\parensmathoper{\mathit{right}}}
\newcommand{\phileft}{\phi_{\mathit{left}}}
\newcommand{\phiright}{\phi_{\mathit{right}}}

\newcommand{\triggerformula}[2]{\mathit{trigger}\ifempty{#1}{}{(#1,#2)}}
\newcommand{\notriggersformula}[2]{\mathit{noTriggers}\ifempty{#1}{}{(#1,#2)}}
\newcommand{\phitrigger}{\phi_{\mathit{trigger}}}
\newcommand{\phinotriggers}{\phi_{\mathit{noTriggers}}}
\newcommand{\phicoreformula}{\phi_{\mathit{core}}}
\newcommand{\phibaseform}{\phi_{\mathit{base}}}
\newcommand{\phibaseformlast}{\phi_{\mathit{baseLast}}}

\newcommand{\baseform}[5]{\Phi_{\mathit{base}}\ifempty{#1}{}{(#1,#2,#3,#4,#5)}}
\newcommand{\baseformlast}[4]{\Phi_{\mathit{baseLast}}\ifempty{#1}{}{(#1,#2,#3,#4)}}
\newcommand{\genform}[1]{\Phi\ifempty{#1}{}{(#1)}}
\newcommand{\coreformula}[4]{\Phi_{\mathit{core}}\ifempty{#1}{}{(#1,#2,#3,#4)}}
\newcommand{\ftp}{\mathit{ftp}}
\newcommand{\fim}{\parensmathoper{\mathit{fim}}}
\newcommand{\limo}{\parensmathoper{\mathit{lim}}}
\newcommand{\fnim}{\parensmathoper{\mathit{fnim}}}
\newcommand{\lnim}{\parensmathoper{\mathit{lnim}}}
\newcommand{\ffim}{\parensmathoper{\mathit{ffim}}}
\newcommand{\flim}{\parensmathoper{\mathit{flim}}}

\newcommand{\intervalDom}{\mathbb{I}}
\newcommand{\req}{r}
\newcommand{\negatetime}{\parensmathoper{\mathit{dual}}}
\newcommand{\fretsem}[1]{\mathcal{F}\BBrackets{#1}}

\copyrightyear{2022}
\acmYear{2022}
\setcopyright{licensedusgovmixed}
\acmConference[CPP '22]{Proceedings of the 11th ACM SIGPLAN International Conference on Certified Programs and Proofs}{January 17--18, 2022}{Philadelphia, PA, USA}
\acmBooktitle{Proceedings of the 11th ACM SIGPLAN International Conference on Certified Programs and Proofs (CPP '22), January 17--18, 2022, Philadelphia, PA, USA}
\acmPrice{15.00}
\acmDOI{10.1145/3497775.3503685}
\acmISBN{978-1-4503-9182-5/22/01}

\begin{document}

\title{A Compositional Proof Framework for \fretish{} Requirements}

\author{Esther Conrad}
\affiliation{
  \institution{NASA Langley Research Center}
  \city{Hampton}
  \state{VA}
  \country{USA}
}
\email{esther.d.conrad@nasa.gov}

\author{Laura Titolo}
\affiliation{
  \institution{National Institute of Aerospace}
  \city{Hampton}
  \state{VA}
  \country{USA}
}
\email{laura.titolo@nianet.org}

\author{Dimitra Giannakopoulou}
\affiliation{
  \institution{NASA Ames Research Center}
  \city{Moffett Field}
  \state{CA}
  \country{USA}
}
\email{dimitra.giannakopoulou@nasa.gov}

\author{Thomas Pressburger}
\affiliation{
  \institution{NASA Ames Research Center}
  \city{Moffett Field}
  \state{CA}
  \country{USA}
}
\email{tom.pressburger@nasa.gov}

\author{Aaron Dutle}
\affiliation{
  \institution{NASA Langley Research Center}
  \city{Hampton}
  \state{VA}
  \country{USA}
}
\email{aaron.m.dutle@nasa.gov}

\titlenote
{The authors would like to thank C\'{e}sar A. Mu\~{n}oz for his valuable help during the initial development of the proof framework.\\
Research by Laura Titolo was supported by the
National Aeronautics and Space Administration under NASA/NIA Cooperative Agreement NNL09AA00A.
}

\begin{abstract}
Structured natural languages provide a trade space between
ambiguous natural languages that make up most written requirements,
and mathematical formal specifications such as Linear
Temporal Logic.
\fretish{} is a structured natural language for the elicitation of system requirements developed at NASA.
The related open-source tool \Fret{} provides support
for translating \fretish{} requirements into temporal logic formulas that can be input to several verification and analysis tools.
In the context of safety-critical systems, it is crucial to ensure that a generated formula captures the semantics of the corresponding \fretish{} requirement precisely.
This paper presents a rigorous formalization of the \fretish{} language including a new denotational semantics
and a proof of semantic equivalence between \fretish{} specifications and their temporal logic counterparts computed by \fret.
The complete formalization and the proof have been developed in the Prototype Verification System (PVS) theorem prover.
\end{abstract}



\maketitle

\section{Introduction}
\label{sec:intro}

Natural language requirements are typically ambiguous and not amenable to be input to formal methods tools.
Conversely, formal mathematical notations are unambiguous but they require domain-specific expertise and can be unintuitive and hard to specify.
Structured natural languages provide a good trade-off between natural language and formal mathematical notation.
\fretish{}~\cite{GiannakopoulouP20} is a restricted structured natural language developed at NASA for writing unambiguous requirements.
A \fretish{} requirement is composed by five fields: scope, condition, component, timing, and response.
The Formal Requirements Elicitation Tool (\Fret{})~\cite{GiannakopoulouP20a} provides support for writing specifications in \fretish{} and for generating corresponding metric temporal logic (MTL) formulas that can be input to several formal verification tools.
\Fret{} currently outputs formulas in the language of NuSMV~\cite{CimattiCGR00} and in CoCoSpec~\cite{ChampionGKT16} syntax.
To improve the confidence in the correctness of the generated formula, an extensive automated testing framework has been presented in \cite{GiannakopoulouP21}.
Testing increases the confidence in the tool, but cannot guarantee full coverage of all cases.
In order to use \Fret{} in a safety-critical context, such as the software architecture for UAV presented in \cite{DuttleMCG2020}, it is ideal to formally guarantee that the semantics of the \fretish{} requirement is preserved in the generated temporal logic specification.

This paper presents a rigorous formalization of the \fretish{} language.
This formalization includes a denotational semantics for the \fretish{} language, and a rigorous proof of correctness of the MTL formula generation algorithm implemented in \fret{}.
The denotational semantics maps a \fretish{} requirement into a set of traces where each state is a set of formulas that holds at a certain point in time.
The proof of correctness ensures that a trace belongs to the semantics of a \fretish{} requirement if and only if the trace is a model of the MTL formula generated by \Fret{} for that requirement.
This formalization\footnote{The PVS formalization is available at \url{https://lauratitolo.github.io/}} was carried out in the Prototype Verification System (PVS)~\cite{pvs}.
The correctness proof has been designed to be compositional on the \fretish{} requirement fields.
This is crucial to building a compact and modular set of definitions and theorems that can be easily extended over time as the \fretish{} language evolves with new constructs and features.
Besides providing a robust proof framework for the \fretish{} language, this research effort gave useful insights on how to simplify the MTL generation algorithms and helped improve the simulation and explanation capabilities of \fret.
To the best of the authors' knowledge, this is the first formalization of a structured natural language in a theorem prover.

The paper is organized as follows. 
In \smartref{sec:preliminaries}, MTL and other preliminary notions are presented.
The \fretish{} language is presented in \smartref{sec:fret}.
\smartref{sec:sem} introduces a new denotational semantics for \fretish{}.
In \smartref{sec:translation}, the algorithm used by \fret{} to generate an MTL formula from a \fretish{} requirement is presented.
\smartref{sec:proving} shows the main results on the correctness of this algorithm.
\smartref{sec:lesson} illustrates the advantages of the proposed formalization.
Related work is discussed in \smartref{sec:related}.
\smartref{sec:conclusion} concludes the paper. 
\section{Metric Temporal Logic}
\label{sec:preliminaries}

Metric Temporal Logic (MTL)~\cite{Koymans90} is an extension of Linear Temporal Logic (LTL) in which the temporal operators are augmented with timing constraints.
In this paper, the past-time fragment of MTL is considered.
This choice is guided by the ultimate goal of the authors of using this formalization to ensure the correctness of runtime monitors automatically generated for autonomous systems, as described in \cite{DuttleMCG2020}.
In fact, past-time temporal logic is usually preferred to its future-time counterpart in the formalization of runtime monitors.

Past-time formulas look at the portion of the execution that has occurred up to the state where they are interpreted.
A past-time formula is satisfied by an execution trace if the formula holds at the final state of the trace.

Let $\intervalDom$ denote the set of intervals of natural numbers of the form $[l,u]$ such that $l,u\in\N$, $l\leq u$, and for all $x\in\N$, $x\in [l,u] \iff l \leq x\leq u$. 
Given a set $\Psyms$ of atomic formulas, the set of past-time MTL formulas is generated by the following grammar.
\begin{align*}
\phi ::= & \true \mid \false \mid p
\mid \neg \phi
        \mid \phi \wedge \phi
        \mid \phi \vee \phi
        \mid \phi \ra \phi
        \\
        & \mid \previous{\phi} 
        \mid \once{\phi}
        \mid \historically{\phi}
        \mid \since{\phi}{\phi}
\end{align*}
where $p\in\Psyms$ and $I\in\intervalDom$. 
When $I$ is omitted it is considered to be $[0,+\infty)$.
The Boolean constants $\true$ and $\false$, and the Boolean connectives $\wedge$, $\vee$, $\ra$, and $\neg$ have the usual logic meaning.
The past-time temporal operators allowed are previous ($\previous{}$), once ($\once[]{}$), historically ($\historically[]{}$), and since ($\since[]{}{}$).
The set of formulas generated with this grammar is denoted with $\MTLDom$.

Let $\vdash$ be the entailment relation between two atomic formulas with respect to a given theory. For instance, if the theory of real numbers arithmetics is considered, it holds that $x-1>0 \vdash x>1$.
The semantics of a past-time MTL formula is given in terms of a satisfaction relation $\models$.
Let $\TraceDom$ be the domain of finite traces of the form $\rho_0 \dots \rho_n$ where for each $i\in[0,n]$, $\rho_i\subseteq\Psyms$.
Given $\rho\in\TraceDom$ and $t\in\N$, $\rho \models_t \phi$ denotes that $\phi$ holds in $\rho$ at time $t$.
\begin{align*}
&\rho \models_t p \iff \rho_t\vdash p\\
&\rho \models_t \neg\phi \iff \rho \not\models_t \phi\\
&\rho \models_t \phi_1 \wedge \phi_2 \iff \rho \models_t \phi_1 \text{, and }\rho \models_t \phi_2\\
&\rho \models_t \phi_1 \vee \phi_2 \iff \rho \models_t \phi_1 \text{ or }\rho \models_t \phi_2\\
&\rho \models_t \phi_1 \ra \phi_2 \iff \rho \not\models_t \phi_1 \text{ or }\rho \models_t \phi_2\\
&\rho \models_t \previous{\phi} \iff \rho \models_{t-1} \phi\, \text{and}\, t\neq 0\\
&\rho \models_t \once[I]{\phi} \iff \exists t_0\in\N:
 t_0 \leq t,\, t-t_0 \in I\, \text{and}\, \rho \models_{t_0} \phi\\
&\rho \models_t \historically[I]{\phi} \iff \forall t_0\in\N:
 t_0 \leq t \wedge\! t-t_0 \in I\, \text{implies}\, \rho \models_{t_0} \phi\\
&\rho \models_t \since{\phi_1}{\phi_2} \iff  
\begin{aligned}[t]
&\exists t_0\in\N:
t_0 \leq t,\, t-t_0 \in I,\,\rho \models_{t_0} \phi_2
 \\&\text{and}\, \forall t_1\in\N\ \text{s.t. } t_0 < t_1 \leq |\rho|: \rho \models_{t_1} \phi_1
 \end{aligned}
\end{align*}
Additionally, $\rho\models\phi$ if $\phi$ holds at the end of the trace $\rho$, \ie{} $\rho\models_{n}\phi$.

Given $m\in\N^{>0}$, the operator $\previous{}^m$ checks if a formula is satisfied exactly $m$ steps before the current time. It is defined as $\previous{}^m \phi = \once[{[m,m]}]{\phi}$.
The operators since-inclusive-required ($\sir{}{}$) and since-inclusive-optional ($\sio{}{}$), introduced in \cite{BauerL11}, are used as syntactic sugar as follows:
\begin{align*}
&\sir{\phi_1}{\phi_2} = \since[]{\phi_1}{(\phi_1 \wedge \phi_2)}\\
&\sio{\phi_1}{\phi_2} = \once[]{\phi_2} \ra \since[]{\phi_1}{(\phi_1 \wedge \phi_2)}.
\end{align*}
Intuitively, $\sir{\phi_1}{\phi_2}$ means that $\phi_1$ needs to hold since and including the point in which $\phi_2$ occurs and moreover $\phi_2$ must eventually occur, while in the optional case $\sio{\phi_1}{\phi_2}$ the formula is only evaluated if $\phi_2$ once holds.

\section{The Fretish Specification Language}
\label{sec:fret}

This section provides an introduction to the \fretish{} language.
More details can be found in \cite{GiannakopoulouP20a,GiannakopoulouP20,GiannakopoulouP21}.

\fretish{} is a restricted natural language for expressing unambiguous requirements.
It aims at providing a natural vocabulary to the user and, at the same time, avoiding ambiguities to ease the translation of the specification into a rigorous mathematical language.

\begin{example}
\label{ex:req1}
Consider the following detect-and-avoid requirement written in natural language: 
 \lq\lq While in flight, a warning alert must be raised within 3 seconds of entering the warning hazard zone of an intruder (250 feet horizontal and 50 feet vertical)\rq\rq.

This requirement can be rephrased in \fretish{} as follows:
\lq\lq In \textit{flight} mode, when $\mathit{horizontal\_distance} \leq 250$
\& $\mathit{vertical\_distance} \leq 50$ the aircraft shall $\fretwithin{}$ 3 seconds satisfy
$\mathit{warning\_alert}$\rq\rq.
\end{example}

\begin{figure*}[tp]
\includegraphics[width=0.95\textwidth]{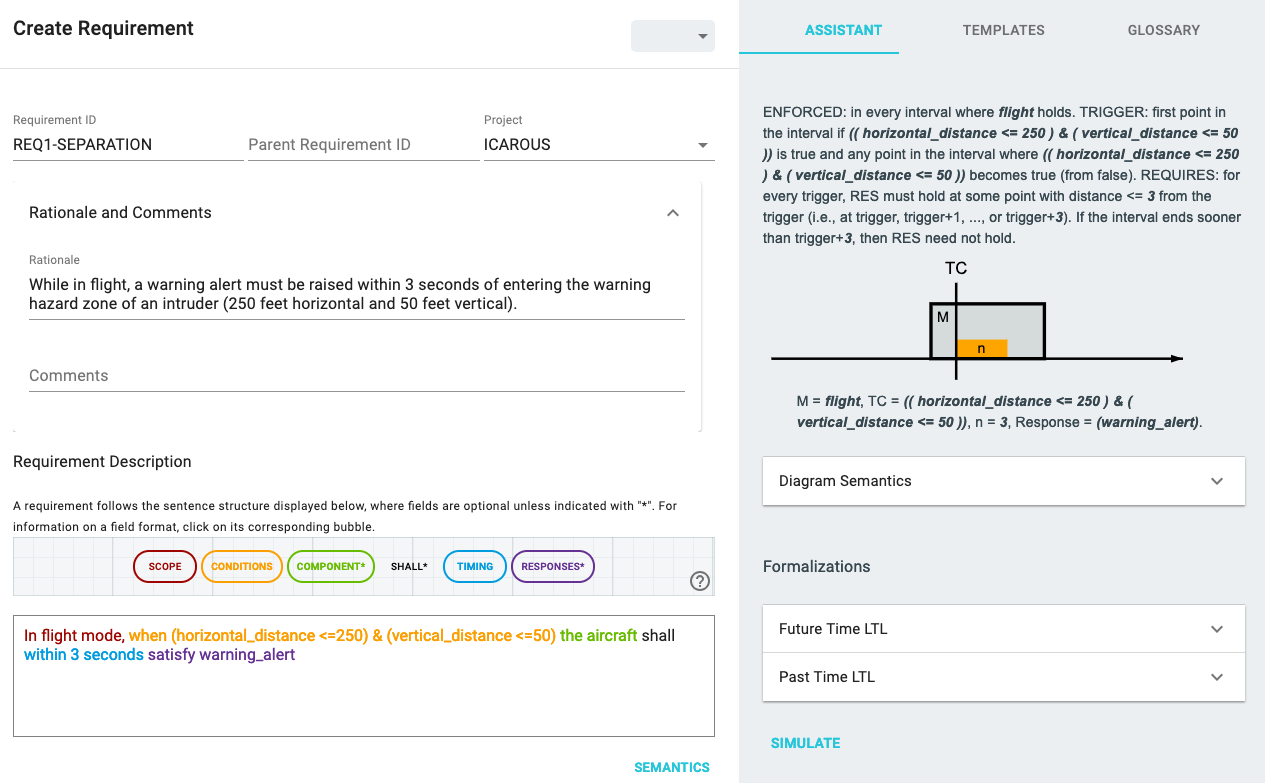}
\caption{\fret{} user interface for the requirement of \smartref{ex:req1}.}\label{fig:gui}
\end{figure*}

The \fret{}~\cite{GiannakopoulouP20a} tool\footnote{\fret{} is available at \url{https://github.com/NASA-SW-VnV/fret}.} parses a \fretish{} requirement, maps it into a template, and translates it into an MTL formula ready to be input to a verification/analysis tool.
\smartref{fig:gui}, depicts the FRET’s requirements elicitation interface for the requirement of
\smartref{ex:req1}.
Once a requirement is entered, the “Semantics” pane shows a text description of the FRETISH requirement, displays a “semantic diagram” showing a visual explanation of the requirement applicability over time, and provides translations from FRETISH to future- and past-time Metric Linear Temporal Logic.

A \fretish{} requirement is parsed into five different fields: \fieldscope{}, \fieldcondition{}, \fieldcomponent{}, \fieldtiming{}, and \fieldresponse{}, three of which are optional: \fieldscope{}, \fieldcondition{}, and \fieldtiming{}.
In addition, the \fieldshall{} keyword must appear and states that the component behavior must conform to the requirement.

The \fieldcomponent{} field specifies the component that the requirement applies to (\eg{}, \emph{``aircraft''} in \smartref{ex:req1}).
The \fieldresponse{} field is of the form \fieldsatisfy{} $\phi$, where $\phi$ is a non-temporal Boolean-valued expression (\eg{}, $\mathit{warning\_alert}$).

Field \fieldscope{} specifies the interval(s) within which the requirement must hold (\eg{}, when in $\textit{flight}$ mode in \smartref{ex:req1}). If the scope is omitted, the requirement is enforced on the entire execution, known as \emph{global} scope.
Given a mode $\mode$, a \fretish{} \fieldscope{} is one of the following relationships: $\fretbefore{\mode}$, $\fretafter{\mode}$, $\fretin{\mode}$, $\fretonlyafter{\mode}$,\\ $\fretnotin{\mode}$, $\fretonlybefore{\mode}$, and $\fretonlyin{\mode}$.

Scope $\fretbefore{\mode}$ indicates that the requirement is enforced strictly before the first point in which $\mode$ holds, $\fretafter{\mode}$ means it is enforced strictly after the last point in which $\mode$ holds, and
$\fretin{\mode}$ that it is enforced while the component is in mode $\mode$.
Scope $\fretnotin{\mode}$ is the dual of $\fretin{\mode}$.
It is sometimes necessary to specify that a requirement can be satisfied \emph{only} in some time frame, meaning it should \emph{not} be satisfied outside of that frame.
For this, the scopes $\fretonlyafter{}$, $\fretonlybefore{}$, and $\fretonlyin{}$ are provided; these will be referred to as {\textit{only} scopes.

Field  \fieldcondition{} is a Boolean expression that triggers the need for a response within the specified scope.
Boolean expressions, familiar to most developers, are used to concisely capture conditions. The set of (non-temporal) Boolean expressions is denoted by $\B$.
If the \fieldcondition{} field is omitted, it is said to be $\nullcond$ and it is equivalent to $true$. Therefore, the trigger is the beginning of the trace.
For instance, the condition in \smartref{ex:req1} is $\mathit{horizontal\_distance}\leq 250$ \& $\mathit{vertical\_distance}\leq 50$.
\smartref{fig:scopes} illustrates how the different scopes are defined for a given mode.
\begin{figure}[tp]
\includegraphics[width=0.48\textwidth]{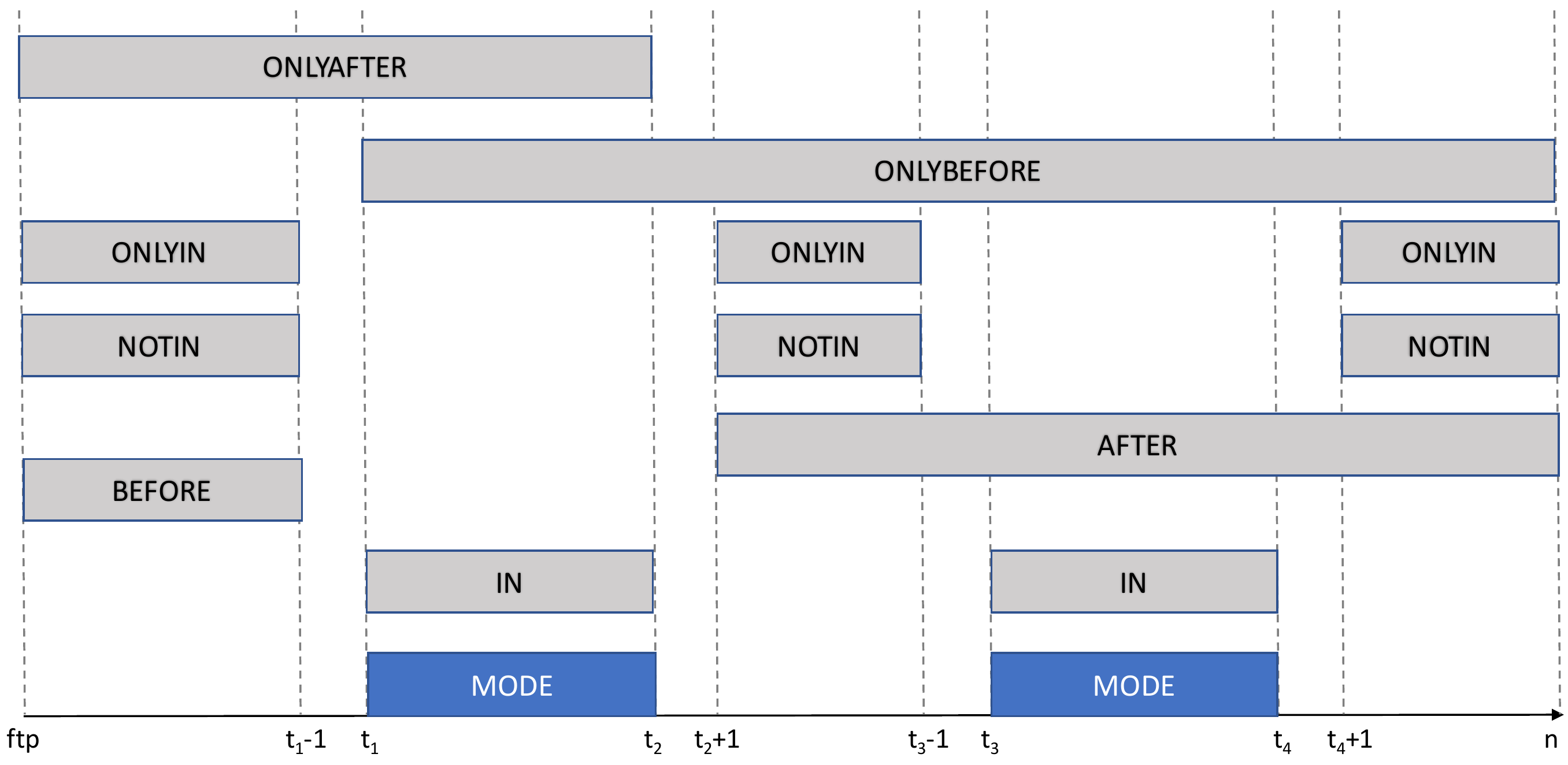}
\caption{Scope intervals definition.}\label{fig:scopes}
\end{figure}

Field \fieldtiming{} specifies when the response is expected relative to each trigger (\eg{}, $\fretwithin{}$ 3 seconds in \smartref{ex:req1}).
Given a duration $d\in\N$, and a stop condition $\stopcond\in\B$, there are nine possibilities for the timing field:
$\fretimmediately$,
in the $\fretnext$ time unit,
$\freteventually$,
$\fretalways$,
$\fretnever$,
$\fretwithin{}$ $d$ time units,
$\fretfor{}$ $d$ time units,
$\fretuntil{}$ the specified stop condition $\stopcond$ occurs,
$\fretbefore{}$ the specified stop condition $\stopcond$ occurs,
and $\fretafter{}$ $d$ time units (interpreted as not for $d$ time units and at the $d+1$ time unit).
When timing is omitted, it is assumed to be $\freteventually$.

A \fretish{} requirement is characterized by the tulple of its fields \fieldscope{}, \fieldcondition{}, \fieldtiming{}, and \fieldresponse{}.
The field \fieldcomponent{} is not relevant to the semantics of the requirement, therefore it is omitted in the definition of requirement.
\begin{definition}[\fretish{} requirement]
    Given $d\in\N$ and $\mode,\stopcond,\res\in\B$, a \emph{\fretish{} requirement} is a tuple $\fretReq{\fieldscope}{\fieldtiming}{\cond}{\res}$
    where
    \begin{align*}
    \fieldscope&\in\{\fretnull,\fretin{\mode},\fretnotin{\mode},\fretbefore{\mode},\\
    &\fretafter{\mode},\fretonlyafter{\mode},\\
    &\fretonlybefore{\mode},\fretonlyin{\mode}\}\\
    \fieldcondition&\in\B\cup\{\nullcond\}\\
    \fieldtiming&\in\{\fretimmediately,\fretnext,\fretnever,\freteventually,\fretalways,\\
    &\fretwithin{d}, \fretfor{d}, \fretaftertime{d},\\
    &\fretuntil{\stopcond}, \fretbefore{\stopcond}\}.
    \end{align*}
\end{definition}

\section{A Denotational Semantics for Fretish}
\label{sec:sem}

In \cite{GiannakopoulouP21}, the semantics of \fretish{} is described in terms of a discrete fragment of the Real-Time Graphical Interval Logic (RTGIL)~\cite{MoserMRKD96}.
RTGIL is a graphical language that interprets linear-time temporal formulas over ordered lists of intervals.
The RTGIL toolset includes a graphical editor and a formula satisfiability checker implemented using tableaux.
However, this checker is no longer maintained.

The RTGIL semantics provides an intuitive graphical representation which is helpful to visualize the meaning of the requirement.
In order to obtain a compositional theoretical framework, this section introduces a denotational semantics for \fretish{} that mimics the RTGIL semantics of \cite{GiannakopoulouP21}.
As part of the work presented in this paper, this semantics has been formalized in the \PVS{} specification language, which offers the necessary expressive power and the possibility to rigorously reason about the properties of the semantics via the \PVS{} theorem prover.

The proposed semantics is based on the notion of \emph{ordered list of intervals}.
\begin{definition}
An ordered list of intervals ($\oli$) $l$ is a finite list of closed intervals of natural numbers in $\intervalDom$ of the form $l = \langle[a_0,b_0],\dots,[a_k,b_k]\rangle$ where $k\in\N$ such that for all $0 \leq i \leq k$, $a_i\leq b_i$, $b_i < a_{i+1}-1$, $a_0 \geq 0$.
  The empty $\oli$ is denoted as $\emptylin$.
\end{definition}
The $i$-th interval $[a_i,b_i]$ in $l$ is denoted by $l_i$.
By abuse of notation, given $x\in\N$, $x\in l$ denotes that there exists $i$ such that $x\in l_i$.
The size of $l$ is denoted as $|l|$.
The lower bound $l_0$ of $l$ is denoted with $\lb{l}$, while the upper bound $b_k$ is denoted with $\ub{l}$.
The same notation is used for denoting the lower and upper bound of a single interval, \ie{} $\lb{[a_i,b_i]}=a_i$ and $\ub{[a_i,b_i]}=b_i$.

An $\oli$ $l$ is said to be bounded by $n\in\N$ if $\ub{l}\leq n$.
The set of all $\oli$ bounded by $n$ is denoted by $\LNI{n}$, or $\LNI{}$ when $n$ is clear from the context or irrelevant.
\begin{figure}[tp]
\includegraphics[width=0.48\textwidth]{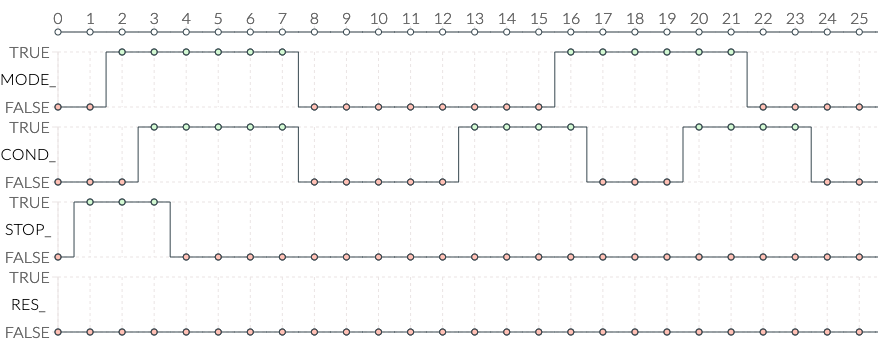}
\caption{An example trace with the four abstract properties of \fretish{} requirements.}\label{fig:OLIs}
\end{figure}

An $\oli$ encodes when a certain property holds along a trace of states indexed by natural numbers.
For \fretish{} requirements, these properties consist of Boolean expressions $\mode$, $\stopcond$, $\cond$, and $\res$.
Notice that in \fretish{} time is discrete. For this reason, the domain of $\oli$s and the traces are indexed by natural numbers, not reals.
While continuous time may be a more accurate representation in some situations, it introduces difficulties in operators like \lq\lq next\rq\rq{} and would require a notion of computational step to be included in execution traces.

Given a Boolean expression $\psi$ and a trace $\rho$, the function $\boolSem{}{}$ computes the $\oli$ that encodes when $\psi$ holds in $\rho$.
In the example trace $\rho$ of Figure~\ref{fig:OLIs}, $\boolSem{mode}{\rho}  = \langle[2,7],[16,21]\rangle$, and $\boolSem{res}{\rho} = \emptylin$.
The function $\boolSem{}{}$ is defined formally below.
\begin{definition}
\label{def:oliexpr}
  Given $\psi\in\B$ and $\rho\in\TraceDom$,
  $\boolSem{\psi}{\rho} \in \LNI{}$ is such that
$i\in \boolSem{\psi}{\rho} \iff \rho \models_i \psi$.
\end{definition}

The notion of complement models when a Boolean expression does not hold, \ie{} its negation holds.
\begin{definition}
\label{def:olicomp}
  Let $l\in\LNI{}$, the \emph{complement} $\bar{l}\in\LNI{}$ of $l$
is such that, for every $x\in\N$, $x\leq n$, $x \notin l \iff x \in \bar{l}$.
\end{definition}

The following lemma follows directly from \smartref{def:oliexpr} and \smartref{def:olicomp}.
\begin{lemma}
\label{lem:olicomp}
Given $\psi\in\B$, $\rho\in\TraceDom$, and $i\in\N$:
$$i\in \boolSem{\psi}{\rho} \iff i\in \overline{\boolSem{\neg \psi}{\rho}}.$$
\end{lemma}

In the following, the \fretish{} semantics is described compositionally, through the values of fields \emph{scope}, \emph{condition}, and \emph{timing}}.
The \emph{scope semantics} defines the list of intervals within which the temporal requirement specification must hold.
\begin{definition}[Scope Semantics]
Let $\rho\in\TraceDom$ such that $|\rho|>0$, $n=|\rho|-1$ and $\modeInt = \boolSem{\mode}{\rho}$.
The semantics of a scope $s$ is defined as follows.
\begin{align*}
    &\scopeSem{\fretnull{}} = \langle[0,n]\rangle
    \\
    &\scopeSem{\fretin{\mode}} = \modeInt
    \\
    &\scopeSem{\fretnotin{\mode}} = \neglni{\modeInt}
    \\
    &\scopeSem{\fretafter{\mode}} =
       \begin{cases}
       \emptylin \qquad\qquad\quad \begin{aligned}
                       &\text{if }\modeInt = \emptylin\\
                       & \text{or }\ub{\modeInt_0} = n
                       \end{aligned}\\
       \langle[\ub{\modeInt_0} + 1, n]\rangle \quad \text{otherwise}
       \end{cases}
    \\
    &\scopeSem{\fretbefore{\mode}} =
       \begin{cases}
       \emptylin  \qquad \qquad\qquad \lb{\modeInt_0} = 0\\
       \langle[0, \lb{\modeInt_0} - 1]\rangle \quad \text{otherwise}
       \end{cases}
    \\
    &\scopeSem{\fretonlyin{\mode}} = \neglni{\modeInt}
    \\
    &\scopeSem{\fretonlyafter{\mode}} =
       \begin{cases}
       \langle[0,n]\rangle \qquad\qquad \text{if $\modeInt = \emptylin$}\\
       \langle[0,\ub{\modeInt_0}]\rangle \quad \text{otherwise}
       \end{cases}
    \\
    &\scopeSem{\fretonlybefore{\mode}} =
       \begin{cases}
       \emptylin \qquad\qquad\qquad\,\, \text{if $\modeInt = \emptylin$}\\
       \langle[\lb{\modeInt_0},n]\rangle \quad\text{otherwise}
       \end{cases}
\end{align*}
\end{definition} 
As already mentioned, note that scopes of type \fretonly{}, mandate that a requirement does \emph{not} hold outside of their corresponding scope.
Thus, the semantics of an only scope is defined as the complement of the semantics of its corresponding regular scope.
Within an interval $I$, a requirement is triggered at each index of $I$, where the conditional expression $\cond$ becomes true from false in $\rho$, and at $\lb{I}$, if $\rho \models_{\lb{I}} \cond$. 
When no condition is specified ($\nullcond$), the requirement is only triggered at $\lb{I}$.
For a trace $\rho$, an interval $I$, and a condition $\cond$, the function $\triggers{\cond{}}{\rho}{I}$ returns the set of all such indices. 
\begin{definition}[Triggers]
Let $\cond{}\in\B\cup\{\fretnull\}$, $I\in\intervalDom$, and $\rho\in\TraceDom$, $\triggers{\cond{}}{\rho}{I}\subseteq \N$ is defined as follows.
\begin{align*}
&\triggers{\cond{}}{\rho}{I} =
\{\lb{I} \mid \cond{} = \fretnull\} \cup{}\\
&\quad\bigcup_{0\leq j < |\condInt|}
\{\mathit{max}(\lb{\condInt_j},\lb{I}) \mid
\begin{aligned}[t]
&\cond{} \neq \fretnull,\\
&\condInt_j\cap I \neq\emptyset\}
\end{aligned}
\end{align*}
\end{definition}
where $\condInt = \boolSem{\cond}{\rho}$.

When, within the range of interval $I$, a condition is always false in $\rho$, the function $\triggers{}{}{}$ will return an empty set; the requirement is never triggered and is therefore vacuously true.
Note that this is different from the case where the condition field in the \fretish{} requirement is $\fretnull$ and, thus, the trigger occurs at $\lb{I}$.

Stop conditions, necessary for the timings $\mathit{until}$ and $\mathit{before}$, are computed similarly to triggers.
Given a stop condition $\stopcond$, the function $\stops{\stopcond}{\rho}{I}$ computes the set of indices where $\stopcond$ is satisfied, within a trace $\rho$, restricted to a particular scope interval $I$.
The function $\firststop{\stopcond}{t}{\rho}{I}$ returns the index of the first occurrence of $\stopcond$ after a trigger $t$, or $\ub{I}$, if $\stopcond$ is never true within interval $I$.
\begin{definition}[Stops]
Given $\stopcond\in\B$, $t\in\N$, $\rho\in\TraceDom$, and $I\in\intervalDom$, the functions $\stops{}{}{}$ and $\firststop{}{}{}{}$ are defined as follows.
  \begin{align*}
    &\stops{\stopcond}{\rho}{I} =\\
    &\quad {\bigcup_{0\leq j < |stopInt|}}
    \{\mathit{max}(\lb{\stopInt_j},\lb{I}) \mid \stopInt_j\cap I \neq\emptyset\}\\
    &\firststop{\stopcond}{t}{\rho}{I} =
    \begin{cases}
       \ub{I}+1 \qquad\,\, \text{if $\nexists s\in\stops{}{}{}:\ t<s$}\\
\mathit{min}(\{ s\! \in\! \stops{}{}{} \mid  t<s \}) \quad \text{otherwise}
    \end{cases}
  \end{align*}
   where $\stopInt = \boolSem{\stopcond}{\rho}$ and $\stops{}{}{} = \stops{\stopcond}{\rho}{I}$.
\end{definition}

Given a trace $\rho$, a condition $\cond$ and a response $\res$, the semantics of a timing field is the set of intervals of indices $I$ in $\rho$ such that, for each trigger defined by $\cond$, the response $\res$ is satisfied per the particular timing field in $I$. 
\begin{definition}[Timing Semantics]
  Let $n\in\N^{>0}$, $\rho\in\TraceDom$, and $\resInt = \boolSem{\res}{\rho}$, the semantics of a timing field is defined as in \smartref{fig:timingsem}.
\end{definition}
The semantics of timing $\fretimmediately$ is composed of all the intervals $I$ such that the triggers occurring in $I$ are also in the response.
Similarly, the semantics of timing $\fretnext$ contains all the intervals $I$ such that the time index following a trigger is in the response.

For timing $\fretalways$, the semantics contains all the intervals $I$ such that all the indices ranging from the first trigger that occurs until the end of the interval are also contained in the response.
The semantics of timing $\fretnever$ is defined as the semantics of timing $\fretalways$ with the negated response.
From \smartref{lem:olicomp}, this is equivalent to checking that all the indices from the first trigger until the end of the interval are included in the complement of the response.

The semantics of the $\freteventually$ timing includes all the intervals $I$ such that there exists an index which is both a trigger and it is included in the interval from the last trigger to the end of the interval.
It is worth noting that, for each trigger $t$, the response is required to hold at least once in the interval from $t$ until the end of the interval.
Therefore, it is sufficient to check that it holds from the last trigger until the interval upper bound.
Similarly, for the $\fretalways$ timing it is sufficient to check every index from the first trigger until the interval upper bound.

The semantics of the $\fretwithin{d}$ timing contains all the intervals $I$ such that, for all the triggers $t$, either there are less than $d$ time instants between $t$ and the end of the interval ($t+d\not\in I$), or there exists a $k$ such that $t+k$ is in the interval $I$ and the response holds at $t+k$.

In the case of the $\fretfor{d}$ timing, the semantics includes all the intervals $I$ such that for all the natural numbers $k$ between 0 and $d$, either $t+k$ is not included in $I$ or $t+k$ is in the response.

The $\fretuntil{\stopcond}$ semantics collects all the intervals such that, for each trigger $t$, the subinterval ranging from $t$ to the first instant satisfying the stop condition $\stopcond$ is included in the response.

The semantics of timing $\fretbefore{\stopcond}$ includes all the intervals that do not contain at least one stop condition
and the intervals such that a stop condition never occurs after a trigger ($\firststop{\stopcond}{t}{\rho}{I} = \ub{I} + 1$) . In these cases the requirement trivially holds. In addition, it collects the intervals such that at least a trigger $t$ exists, $\stopcond$ holds between $t$ and the end of the interval, and there exists a time instant between the trigger and the stop condition that satisfies the response.

Finally, timing $\fretafter{}$ semantics is derived from the semantics of $\fretfor{}$ and $\fretwithin{}$. In fact, $\fretafter{}$ $d$ time units is equivalent to not $\fretfor{}$ $d$ time units and $\fretwithin{}$ $d+1$ time units.
In addition, for all the timings, the semantics includes all the intervals that do not contain any trigger. This reflects the fact that when no trigger occurs, the requirement holds trivially.
\begin{figure*}
\centering{
\begin{align*}
    &\timeSem{\fretimmediately}{\cond}{\res} =
    \begin{aligned}[t]
    &\{I \mid
    \triggers{\cond{}}{\rho}{I}=\emptyset\}\\
    &\cup\{I \mid \triggers{\cond{}}{\rho}{I}\neq\emptyset
    \wedge \forall t\in \triggers{\cond{}}{\rho}{I}:\ t\in\resInt \}
    \end{aligned}
    \\[1ex]
&\timeSem{\fretnext}{\cond}{\res} =
    \begin{aligned}[t]
    &\{I \mid \triggers{\cond{}}{\rho}{I}=\emptyset\}\\
    &\cup\{I \mid
    \triggers{\cond{}}{\rho}{I} \neq\emptyset \wedge \forall t \in \triggers{\cond{}}{\rho}{I}:\
    t+1\in I \ra t+1\in\resInt \}
    \end{aligned}
    \\[1ex]
&\timeSem{\fretalways}{\cond}{\res} =
    \begin{aligned}[t]
    &\{I  \mid\triggers{\cond{}}{\rho}{I}=\emptyset\}\\
    &\cup\{I \mid
    \triggers{\cond{}}{\rho}{I}\neq\emptyset \wedge{}
    \forall j\in[\mathit{min}(\triggers{\cond{}}{\rho}{I}),\ub{I}]:\ j\in\resInt \}
    \end{aligned}
    \\[1ex]
&\timeSem{\fretnever}{\cond}{\res} = \timeSem{\fretalways}{\cond}{\neg\res}
    \\[1ex]
&\timeSem{\freteventually}{\cond}{\res} =
    \begin{aligned}[t]
    &\{I \mid \triggers{\cond{}}{\rho}{I}=\emptyset\}\\
    &\cup\{I \mid
    \triggers{\cond{}}{\rho}{I}\neq\emptyset \wedge{}
    \exists j\in[\mathit{max}(\triggers{\cond{}}{\rho}{I}),\ub{I}]:\ j\in\resInt \}
    \end{aligned}
    \\[1ex]
&\timeSem{\fretwithin{d}}{\cond}{\res} =
    \begin{aligned}[t]
    &\{I \mid \triggers{\cond{}}{\rho}{I}=\emptyset\}\\
    &\cup\{I \mid
    \begin{aligned}[t]
     &\triggers{\cond{}}{\rho}{I}\neq\emptyset \wedge \forall t\in \triggers{\cond{}}{\rho}{I}:\\
     &t + d \in I \ra \exists k\leq d:\ t+k \in I \wedge t+k \in\resInt \}
    \end{aligned}
    \end{aligned}
    \\[1ex]
&\timeSem{\fretfor{d}}{\cond}{\res} =
    \begin{aligned}[t]
    &\{I \mid \triggers{\cond{}}{\rho}{I}=\emptyset\}\\
    &\cup\{I \mid
    \triggers{\cond{}}{\rho}{I}\neq\emptyset \wedge \forall t\in \triggers{\cond{}}{\rho}{I}
    \forall k\leq d:\ t+k \in I \ra t+k \in\resInt \}
    \end{aligned}
    \\[1ex]
&\timeSem{\fretuntil{\stopcond}}{\cond}{\res} =
    \begin{aligned}[t]
    &\{I \mid \triggers{\cond{}}{\rho}{I}=\emptyset\}\\
&\cup\{I \mid
    \begin{aligned}[t]
     &\triggers{\cond{}}{\rho}{I}\neq\emptyset \wedge \forall t\in \triggers{\cond{}}{\rho}{I}:\\
     &\forall x\in [t,\firststop{\stopcond}{t}{\rho}{I}-1]: x\in\resInt\}
     \end{aligned}
    \end{aligned}
    \\  
    &\timeSem{\fretbefore{\stopcond}}{\cond}{\res} =
    \begin{aligned}[t]
    &\{I \mid \triggers{\cond{}}{\rho}{I}=\emptyset \vee{} \stops{\stopcond}{\rho}{I}=\emptyset\}\\
&\cup{}\{I \mid
    \begin{aligned}[t]
    &\triggers{\cond{}}{\rho}{I}\neq\emptyset \wedge \stops{\stopcond}{\rho}{I}\neq\emptyset
    \wedge \forall t\in \triggers{\cond{}}{\rho}{I}:
    \\& 
    \firststop{\stopcond}{t}{\rho}{I} = \ub{I} + 1 \vee{}
    \exists x: t\leq x<\firststop{\stopcond}{t}{\rho}{I} \wedge x\in\resInt\}
    \end{aligned}
    \end{aligned}
    \\[1ex]
    &\timeSem{\fretafter{n}}{\cond}{\res} =
    \timeSem{\fretfor{n}}{\cond}{\neg\res}
    \cap
    \timeSem{\fretwithin{n+1}}{\cond}{\res}
\end{align*}
}
\caption{Timing field semantics.}\label{fig:timingsem}
\end{figure*}

The semantics of a \fretish{} requirement is defined compositionally from the semantics of its fields.
The function $\negatetime{}$ returns the dual of a timing field and it is defined in \smartref{tbl:timingdual}.
\begin{definition}[\fretish{} semantics]
  Let $\req=\langle\fieldscope,$\\
  $\fieldtiming,\cond,\res\rangle$ be a requirement.
  The semantics of $r$ is defined as follows.
  $$\fretsem{\req} =
  \begin{aligned}[t]
  &\{ \rho\in \TraceDom \mid
  \begin{aligned}[t]
  &\scopeSem{\fieldscope} \neq \emptylin \text{ and }\\
  &\forall i \text { such that } 0\leq i < |\scopeSem{\fieldscope}|:\\
  &(\scopeSem{\fieldscope})_i \in \timeSem{\fieldtiming'}{\cond}{\res'}\}\\
  \end{aligned}
\end{aligned}$$
  where
  \begin{align*}
  \fieldtiming' &= \begin{cases}
  \negatetime{\fieldtiming} & \text{if $\fieldscope$ is of type only}\\
  \fieldtiming &\text{otherwise}
  \end{cases}\\
  \res' &= \begin{cases}
  \neglni{\res} & \text{if $\fieldscope$ is of type only}\\
  \res &\text{otherwise}
  \end{cases}
  \end{align*}
\end{definition}
\begin{table}
\caption{Timing duality.}
\label{tbl:timingdual}
\begin{tabular}{ |l|l| } 
\hline\multicolumn{1}{|c|}{$\fieldtiming$}
& \multicolumn{1}{c|}{$\negatetime{\fieldtiming}$}\\
\hline\hline
$\fretalways$ & $\freteventually$ \\
\hline
$\freteventually$ & $\fretalways$ \\
\hline
$\fretwithin{d}$ & $\fretfor{d}$ \\
\hline
$\fretfor{d}$ & $\fretwithin{d}$ \\
\hline
$\fretbefore{\stopcond}$ & $\fretuntil{\stopcond}$ \\
\hline
$\fretuntil{\stopcond}$ & $\fretbefore{\stopcond}$ \\
\hline
$\fretimmediately$ & $\fretimmediately$ \\
\hline
$\fretnext$ & $\fretnext$ \\
\hline
\end{tabular}
\end{table}

As already mentioned, for scopes of type \fretonly{}, a requirement does \emph{not} hold outside of their corresponding scope.
This means that the negation of the main body of the requirement must hold outside their scope.
Therefore, the dual of the timing is considered and the result is negated.

The \PVS{} formalization of \fretish{} includes an executable implementation of this semantics and theorems ensuring that it is equivalent to the denotational semantics presented in this section.
Additionally, to improve the confidence that the \PVS{} implementation is equivalent to the RTGIL semantics implemented in \fret{}, the proposed semantics has been checked with the help of the \fret{} extensive testing framework.
To this aim, the testing framework has been enhanced to automatically generate test cases in the language of \PVS{}.
In particular, 10,000 test cases have been generated. 
Each of these test cases was composed of a $\fretish$ specification $r$, a trace $\rho$, and a Boolean indicating if the trace belongs to the semantics of the specification.
This Boolean value was computed by a component of \fret{}, called \emph{oracle}, which implements the RTGIL semantics presented in \cite{GiannakopoulouP21}.
Finally, the executable \PVS{} semantics was executed for each test case, and a Boolean value was computed indicating if the trace $\rho$ was included in the semantics of $r$.

\section{From \fretish{} Specifications to \mtl{} Formulas}
\label{sec:translation}
The \fret{} tool generates a temporal logic formula from a \fretish{} specification.
While the tool supports both future and past time metric linear temporal logic, in this work the past time fragment is considered.
The formula is built compositionally on the \fretish{} requirement fields.
The algorithm views a trace as a collection of disjoint intervals characterized by the scope field.
The temporal requirement, characterized by the timing and condition fields, is imposed on each of these intervals.

First, the formula corresponding to the timing field and condition is generated. This formula is called \emph{core formula}.
The following functions are used to model the condition field.
Given a condition $\cond\in\B$ and a formula $\phileft\in\B$ that models the left endpoint of the considered interval, the formula $\triggerformula{\cond}{\phileft}$ characterizes a trigger, \ie{} a point in time when $\cond$ becomes true from false or when it holds at the beginning of the interval.
\begin{equation}
\label{eq:triggerformula}
\triggerformula{\cond}{\phileft} = (\cond \wedge \previous{(\neg\cond)}) \vee (\cond \wedge \phileft)
\end{equation}

The formula $\notriggersformula{}{}$ captures the fact that the condition never occurs in the interval.
In this case, the requirement is trivially true.
\begin{equation}
\label{eq:notriggers}
\notriggersformula{\cond}{\phileft} = \sir{\neg\cond}{\phileft}
\end{equation}

Given a timing field $\fieldtiming$, a condition $\cond\in\B$, a response $\res\in\B$, and a left endpoint $\phileft$, $\coreformula{}{}{}{}$ computes the core formula as illustrated in \smartref{tbl:coreformula}.
Here, $\phitrigger$ denotes $\triggerformula{\cond}{\phileft}$, while $\phinotriggers$ denotes the formula $\notriggersformula{\cond}{\phileft}$.
\begin{table*}[tp]
\caption{Core formula definition for $\nullcond$ condition or $\cond\in\B$.}
\label{tbl:coreformula}
\begin{tabular}{ |l|l|l| }
\hline\multicolumn{1}{|c|}{$\fieldtiming$}
& \multicolumn{1}{|c|}{$\coreformula{\fieldtiming}{\nullcond}{\res}{\phileft}$}
& \multicolumn{1}{c|}{$\coreformula{\fieldtiming}{\cond}{\res}{\phileft}$}
\\
\hline\hline
& \\[-2.5ex]
$\fretimmediately$ & $\phileft \ra \res$ & $\phitrigger \ra \res$\\[0.5ex]
\hline
& \\[-2.5ex]
$\fretnext$ & $(\previous{\phileft}) \ra \res$ & $(\previous{\phitrigger}) \ra (\res \vee \phileft)$\\[0.5ex]
\hline
& \\[-2.5ex]
$\fretalways$  & $\res$ & $\phinotriggers \vee (\sir{\res}{\phitrigger})$\\[0.5ex]
\hline
& \\[-2.5ex]
$\freteventually$ & $\neg (\sir{\neg \res}{\phileft})$ & $\phinotriggers \vee \neg(\sir{\neg\res}{\phitrigger})$\\[0.5ex]
\hline
& \\[-2.5ex]
$\fretuntil{\stopcond}$ & $ (\sir{\neg\stopcond}{\phileft}) \ra \res$ & $\phinotriggers \vee ((\sir{\neg\stopcond}{\phitrigger}) \ra \res)$\\[0.5ex]
\hline
& \\[-2.5ex]
$\fretbefore{\stopcond}$
& $\begin{aligned}[t]\stopcond &\ra (
(\neg\phileft \\ &{}\wedge \neg\previous{(\sir{\neg\res}{\phileft})}))\end{aligned}$
& $\begin{aligned}[t]\stopcond &\ra (\phinotriggers \vee
(\neg\phileft \wedge \neg\phitrigger \\ &{}\wedge \neg\previous{(\sir{\neg\res}{\phitrigger})}))\end{aligned}$\\[1ex]
\hline
& \\[-2.5ex]
$\fretfor{m}$
& ($\once[{[0,m]}] \phileft) \ra \res$
& $(\once[{[0,m]}] \phitrigger) \ra (\phinotriggers \vee \res)$\\[0.5ex]
\hline
& \\[-2.5ex]
$\fretwithin{m}$
& $(\sir{\neg\res}{\phileft}) \ra (\once[{[0,m-1]}] \phileft)$
& $\begin{aligned}[t]\previous^{m}(\phitrigger &\wedge \neg\res)\\ &\ra (\once[{[0,m-1]}] (\phileft \vee \res))\end{aligned}$\\[0.7ex]
\hline
\end{tabular}
\end{table*}

The core formula is then interpreted in a \emph{generic} finite interval defined by two endpoints: $\phileft$, and $\phiright$.
The core formula is checked at each point in the target interval, except for the timing $\freteventually$, which is checked just once at the right endpoint.
The resulting formula is called \emph{base formula}.
\begin{equation}\label{eq:baseform}
\begin{aligned}
&\baseform{\fieldtiming}{\cond}{\res}{\phileft}{\phiright} \\
&\quad=
\begin{cases}
\phiright \ra \previous{\phicoreformula} &\text{if $\fieldtiming = \freteventually$}\\
\phiright \ra \previous{(\sio{\phicoreformula}{\phileft})}&\text{otherwise}
\end{cases}
\end{aligned}
\end{equation}
where $\phicoreformula = \coreformula{\fieldtiming}{\cond}{\res}{\phileft}$.
There is a special case of base formula that is defined when the interval spans to the end of the trace.
In this case, the right endpoint is not defined and the core formula is required to hold from the left endpoint to the end of the trace.
\begin{equation}\label{eq:baseformlast}
\begin{aligned}
&\baseformlast{\fieldtiming}{\cond}{\res}{\phileft} \\
&\quad=
\begin{cases}
(\once[]{\phileft}) \ra \phicoreformula &\text{if $\fieldtiming = \freteventually$}\\
\sio{\phicoreformula}{\phileft}&\text{otherwise}
\end{cases}
\end{aligned}
\end{equation}

Finally, the \emph{general formula} imposes the base formula on the whole execution trace and the generic points $\phileft$ and $\phiright$ are replaced by formulas characterizing the given scope field.
Given $\mode\in\B$, \smartref{tbl:scopeendpoints} shows a list of abbreviations used to model different points of interest in the execution trace.
The abbreviation $\ftp$ denotes the first time point in execution, $\fim{\mode}$ (\resp{} $\limo{\mode}$) is the first (last) state in which $\mode$ holds, $\fnim{\mode}$ ($\lnim{\mode}$) is the first (last) state in which $\mode$ does not hold, and $\ffim{\mode}$ ($\flim{\mode}$) is the first occurrence of $\fim{\mode}$ ($\limo{\mode}$) in the execution.
\begin{table}
\caption{Formulas for point of interest.}
\label{tbl:scopeendpoints}
\begin{tabular}{ |l|l| } 
\hline\multicolumn{1}{|c|}{point}
& \multicolumn{1}{c|}{formula}\\
\hline\hline
$\ftp$ & $\neg\previous{\true}$ \\
\hline
$\fim{\mode}$  & $\mode \wedge (\ftp \vee (\previous(\neg \mode)))$ \\
\hline
$\limo{\mode}$ & $\neg\mode \wedge \previous{\mode}$ \\
\hline
$\fnim{\mode}$ & $\neg\mode \wedge (\ftp \vee (\previous \mode))$ \\
\hline
$\lnim{\mode}$ & $\mode \wedge \previous{(\neg\mode)}$ \\
\hline
$\ffim{\mode}$ & $\fim{\mode} \wedge (\ftp \vee \previous{(\historically[]{\neg\mode})})$ \\
\hline
$\flim{\mode}$ & $\limo{\mode} \wedge \previous{(\historically[]{\neg\limo{\mode}})}$\\
\hline
\end{tabular}
\end{table}

\smartref{tbl:scopeleftright} shows which points of interest in the trace, $\scopeleft{}$ and $\scoperight{}$ correspond to, depending on the $\fieldscope$ field.
It is important to notice that for $\nullscope$, $\afterscope{}$, and $\onlybeforescope{}$ scopes the right endpoint is not defined.
In fact, in these cases, the interval spans until the end of the trace so there is no need to impose a right endpoint.
\begin{table}
\caption{Scope endpoints definition.}
\label{tbl:scopeleftright}
\begin{tabular}{ |l|l|l| } 
\hline\multicolumn{1}{|c|}{$\fieldscope$}
& \multicolumn{1}{c|}{$\scopeleft{\fieldscope}$}
& \multicolumn{1}{c|}{$\scoperight{\fieldscope}$}\\
\hline\hline
$\nullscope$ & $\ftp$ & - \\
\hline
$\beforescope{\mode}$ & $\ftp$ & $\ffim{\mode}$\\
\hline
$\afterscope{\mode}$ & $\flim{\mode}$ & - \\
\hline
$\inscope{\mode}$ & $\fim{\mode}$ & $\limo{\mode}$\\
\hline
$\notinscope{}$/$\onlyinscope{\mode}$ & $\fnim{\mode}$ & $\lnim{\mode}$ \\
\hline
$\onlybeforescope{\mode}$ & $\ffim{\mode}$ & - \\
\hline
$\onlyafterscope{\mode}$ & $\ftp$ & $\flim{\mode}$ \\
\hline
\end{tabular}
\end{table}
The general formula is defined by composing the base formula, depending on the timing and condition fields, and the information about the left and right endpoints, depending on the scope field.
The function $\genform{}{}{}{}$ computes the formula corresponding to a given \fretish{} requirement $\fretReq{\fieldscope}{\fieldtiming}{\cond}{\res}$.
\begin{equation}
\begin{aligned}
&\genform{\fretReq{\fieldscope}{\fieldtiming}{\cond}{\res}}\\
&\qquad=
\begin{cases}
\phibaseformlast \qquad\qquad\quad \text{if $\fieldscope \in
\{\begin{aligned}[t] &\nullscope, \fretafter{}, \fretbefore{}\}\end{aligned}$}\\[1.5ex]
\begin{aligned}
&\historically[]{(\phibaseform \vee \ftp)} \wedge \\
&((\sir{\neg\phiright}{\phileft}) \ra \phibaseformlast)
\end{aligned}
\qquad\quad \text{otherwise}
\end{cases}
\end{aligned}
\end{equation}
where
\begin{align*}
\phileft &= \scopeleft{\fieldscope}\\
\phiright &= \scoperight{\fieldscope}\\
\phibaseformlast &= \baseformlast{\fieldtiming}{\cond}{\res}{\phileft}\\
\phibaseform &= \baseform{\fieldtiming}{\cond}{\res}{\phileft}{\phiright}.
\end{align*}

\begin{example}
Consider the \fretish{} requirement of \smartref{ex:req1}:
\lq\lq In \textit{flight} mode, when $\mathit{horizontal\_distance} \leq 250$
\& $\mathit{vertical\_distance} \leq 50$ the aircraft shall $\fretwithin{}$ 3 seconds satisfy
$\mathit{warning\_alert}$\rq\rq.
The following steps computes the corresponding past-time MTL formula.

First of all, it is necessary to determine the endpoints of each interval in which the requirement is imposed.
Since the scope is $\inscope{\mathit{flight\_mode}}$, the MTL formula characterizing the interval left endpoint is:
$$\fim{\mathit{flight\_mode}}\! =\! \mathit{flight\_mode} \wedge (\ftp \vee (\previous(\neg \mathit{flight\_mode})));$$
while the formula characterizing the right endpoint is:
$$\limo{\mathit{flight\_mode}} = \neg\mathit{flight\_mode} \wedge \previous{\mathit{flight\_mode}}.$$

The condition indicated in the requirement is: 
$$\mathit{horizontal\_distance} \leq 250\! \wedge\! \mathit{vertical\_distance} \leq 50.$$
Thus, a trigger of this condition is modeled by the formula
\begin{align*}
&\phitrigger\! =\!
 (\mathit{horizontal\_distance}\! \leq\! 250 \wedge \mathit{vertical\_distance}\! \leq\! 50 \\
&\quad {} \wedge \previous{(\neg(\mathit{horizontal\_distance} \leq 250)}\\
&\qquad\qquad {}\vee \neg(\mathit{vertical\_distance} \leq 50)))\\
&\quad {} \vee
\begin{aligned}[t](&\mathit{horizontal\_distance} \leq 250 \wedge \mathit{vertical\_distance} \leq 50\\
& \wedge \fim{\mathit{flight\_mode}}).
\end{aligned}
\end{align*}

At this point, it is possible to compute the core formula for the timing $\fretwithin{3}$:
\begin{align*}
\phicoreformula = &\previous^{3}(\phitrigger \wedge \neg\mathit{warning\_alert})\\
 &\quad \ra (\once[{[0,2]}] (\fim{\mathit{flight\_mode}} \vee \mathit{warning\_alert})).
\end{align*}

The regular base formula and the base formula for the last interval case are computed as follows:
\begin{align*}
\label{eq:baseformula}
&\phibaseform = \limo{\mathit{flight\_mode}} \ra \previous{(\sio{\phicoreformula}{\fim{\mathit{flight\_mode}}})}
\\
&\phibaseformlast = \sio{\phicoreformula}{\fim{\mathit{flight\_mode}}}.
\end{align*}
Finally, the general formula that imposes the requirement on the entire trace is:
\begin{equation*}
\label{eq:genformula}
\begin{aligned}
&\historically[]{(\phibaseform \vee \ftp)} \wedge
\\
&((\sir{\neg\limo{\mathit{flight\_mode}}}{\fim{\mathit{flight\_mode}}}) \ra \phibaseformlast).
\end{aligned}
\end{equation*}
\end{example}

\section{Proving the Semantic Equivalence}
\label{sec:proving}
This section presents the main theorems ensuring the correctness of the \fret{} past-time MTL formula generation algorithm. 
As already mentioned, the entire formalization and the proofs have been mechanically checked in the PVS theorem prover.

\fretish{} currently supports 8 relationships for the \fieldscope{} field (including global scope), 2 options for the condition, and 10 options for the timing field, for a total of 160 combinations of semantic templates.
Proving the equivalence of each one of these templates to the corresponding general formula would be extremely time-consuming.
Therefore, to provide a reusable and extensible formalization, the proof has been structured in a compositional manner.

Following the same structure of the formula generation, the first step is to show that the base formula, which depends on the timing field, is equivalent to the semantics of the timing field in the context of an interval $I$.
To this aim, it is convenient to notice that the definition of base formula in Equations~\eqref{eq:baseform} and \eqref{eq:baseformlast}, which use the \emph{since inclusive optional} construct, are equivalent to imposing the core formula on the interval of interest with the historically operator.

\begin{lemma}
\label{lem:baseform_historically}
Let $\rho\in\TraceDom$, $n\in\N$ such that $n = |\rho| -1$, and let $I\in\intervalDom$ and $\phi,\phileft,\phiright\in\MTLDom$ such that
$\rho \models_{\lb{I}} \phileft$, $\rho \models_{\ub{I}+1} \phiright$, and for all $t>\lb{I}$, $\rho \not\models_{t} \phileft$, then if $\ub{I}\neq n$ then
\begin{align*}
&\rho\models_{\ub{I}+1} \phiright \ra \previous{(\sio{\phi}{\phileft})}\\
& \iff \rho\models_{n} \historically[{[n-\ub{I}, n-\lb{I}]}]{\phi},
\end{align*}
otherwise
\begin{align*}
&\rho\models_{\ub{I}} \sio{\phi}{\phileft}
\iff \rho\models_{n} \historically[{[n-\ub{I}, n-\lb{I}]}]{\phi}.
\end{align*}
\end{lemma}
It is worth noting that, since past-time temporal formulas are evaluated starting from the end of the trace at time $n$, when the operator historically ($\historically[]{}$) is used, the formula is imposed on the interval $[n-\ub{I}, n-\lb{I}]$.
This is equivalent to using the future time always (or globally) operator on the interval $I$.
\begin{corollary}
\label{cor:baseform_historically}
Given a timing field $\fieldtiming$, a condition field $\cond\in\B$, a response field $\res\in\B$, a trace $\rho\in\TraceDom$, $n\in\N$ such that $n = |\rho| - 1$, an interval $I\in\intervalDom$, and $\phi,\phileft,\phiright\in\MTLDom$ such that
$\rho \models_{\lb{I}} \phileft$, $\rho \models_{\ub{I}+1} \phiright$, and for all $t>\lb{I}$, $\rho \not\models_{t} \phileft$, if $\ub{I} \neq n$ then
\begin{align*}
&\rho\models_{\ub{I}+1} \baseform{\fieldtiming}{\cond}{\res}{\phileft}{\phiright}\\
&\iff
\rho\models_{n} \historically[{[n-\ub{I}, n-\lb{I}]}]{\coreformula{\fieldtiming}{\cond}{\res}{\phileft}},
\end{align*}
otherwise
\begin{align*}
&\rho\models_{\ub{I}} \baseformlast{\fieldtiming}{\cond}{\res}{\phileft}\\
&\iff
\rho\models_{n} \historically[{[n-\ub{I}, n-\lb{I}]}]{\coreformula{\fieldtiming}{\cond}{\res}{\phileft}}.
\end{align*}
\end{corollary}
The following lemma states that if the $\triggerformula{}{}$ formula defined in Equation~\eqref{eq:triggerformula} is satisfied at an index $t$ of a trace $\rho$, then $t$ belongs to the set of triggers for $\rho$, and vice-versa.
Additionally, if there is no trigger occurring before or at index $t$, then the formula $\notriggersformula{}{}$ is satisfied between the beginning of the scope interval of interest and $t$, and vice-versa.
\begin{lemma}
\label{lem:triggers}
Given $\cond\in\B$, $\rho\in\TraceDom$, $I\in\intervalDom$, and $\phileft\in\MTLDom$ such that $\rho \models_{\lb{I}} \phileft$ and for all $t>\lb{I}$, $\rho \not\models_{t} \phileft$,
for all $t<|\rho|$:
$$t\in\triggers{\cond}{I}{\rho} \iff \rho \models_{t} \triggerformula{\cond}{\phileft}$$
and
$$\begin{aligned}[t]
&\triggers{\cond}{I}{\rho} \cap [\lb{I},t] = \emptyset\\
&\iff \forall t_0\in[\lb{I},t]: \rho\models_{t_0} \notriggersformula{\cond}{\phileft}.
\end{aligned}$$
\end{lemma}
\begin{proof}[Proof Sketch.]
The first equivalence follows from the fact that either $\cond$ is satisfied at $t$ but not at $t-1$ or $\cond$ holds at the beginning of the interval $I$. Thus, by \smartref{def:oliexpr}, $t$ is a lower bound of an interval in $\boolSem{\cond}{\rho}$ or is the interval lower bound $\lb{I}$.
The second equivalence follows from the fact that, by definition of $\sir{}{}$, $\forall t_0\in[\lb{I},t]$,
$\rho\not\models_t \cond$. Thus, $t$ is not included in $\boolSem{\cond}{\rho}$. Vice-versa, if $t$ was a member of $\boolSem{\cond}{\rho}$, then it would satisfy $\cond$, but $\rho\models_t \neg\cond$ by definition of $\sir{}{}$.
\end{proof}
\smartref{tbl:coreformula} distinguishes two cases for the definition of the core formula: condition omitted ($\nullcond$) or condition specified ($\cond\in\B$).
This distinction is convenient since the formulas generated when the condition is $\nullcond$ are more compact.
However, it can be shown that the simplified formula generated for the $\nullcond$ condition is equivalent to the formula generated when the condition is $\true$.
\begin{lemma}
\label{lem:nulltruecond}
Given a timing field $\fieldtiming$, a scope field $\fieldscope$, a condition $\cond\in\B$, a response $\res\in\B$, then
$$\genform{\fretReq{\fieldscope}{\fieldtiming}{\nullcond}{\res}}\!\!\iff\!\!\genform{\fretReq{\fieldscope}{\fieldtiming}{\true}{\res}}.$$
\end{lemma}
\begin{proof}[Proof Sketch.]
From Equations \eqref{eq:triggerformula} and \eqref{eq:notriggers}, observe that $\triggerformula{\true}{\phileft} = \phileft$ and  $\notriggersformula{\true}{\phileft} = \false$.
From \smartref{tbl:coreformula} and Equation~\eqref{eq:genformula}, the result follows by applying Boolean connectives manipulation and simplifications.
\end{proof}
The following lemma states the correctness of the base formula. 
In other words, it shows that the semantics of a timing field is equivalent to the semantics of its corresponding base formula generated as shown in \smartref{sec:translation}.
\begin{theorem}
\label{th:baseformeq}
  Given $\rho\in\TraceDom$, a timing field $\fieldtiming$, a scope field $\fieldscope$, a condition $\cond\in\B$, a response $\res\in\B$,
  an interval $I\in\intervalDom$
  and $\phileft,\phiright\in\MTLDom$ such that $\rho \models_{\lb{I}} \phileft$, $\rho \models_{\ub{I}+1} \phiright$, and for all $t>\lb{I}$, $\rho \not\models_{t} \phileft$, if $\ub{I} \neq n$ then
  \begin{align*}
  &I\in \timeSem{\fieldtiming}{\cond}{\res}\\
  &\qquad\iff
  \rho \models_{\ub{I}+1} \baseform{\cond}{\fieldtiming}{\res}{\phileft}{\phiright}
  \end{align*}
  otherwise
  \begin{align*}
  &I\in \timeSem{\fieldtiming}{\cond}{\res}\\
  &\qquad\iff
  \rho \models_{\ub{I}} \baseformlast{\cond}{\fieldtiming}{\res}{\phileft}.
  \end{align*}
\end{theorem}
The proof proceeds by cases on the timing field.
Below, the case in which $\fieldtiming = \fretalways$ is shown.
The proofs for the other timings are similar and they are available as part of the \fretish{} \PVS{} formalization.
\begin{proof}
Assume that $\fieldtiming = \fretalways$, $\cond\in\B$, and $\ub{I} \neq n$.
From \smartref{lem:nulltruecond}, the case when $\cond=\fretnull$ is equivalent to the case when $\cond=\true$.
Consider the $\implies$ direction.
Let $I \in \timeSem{\fretalways}{\cond}{\res}$ and split the proof into two cases.
Assume, first, that $\triggers{\cond{}}{\rho}{I}=\emptyset$, thus:
$$\begin{aligned}[t]
&\triggers{\cond}{I}{\rho}=\emptyset\\
\implies & \triggers{\cond}{I}{\rho} \cap [\lb{I},\ub{I}] = \emptyset\\
&[\text {By \smartref{lem:triggers}}]\\
\implies & \forall t_0\in[\lb{I},\ub{I}] : \rho \models_{t_0} \phinotriggers\\
\implies & \forall t_0\in[\lb{I},\ub{I}] : \rho \models_{t_0} \phinotriggers \vee (\sir{\res}{\phitrigger})\\
&[\text{By \smartref{tbl:coreformula}}]\\
\implies & \forall t_0\in[\lb{I},\ub{I}] : \rho \models_{t_0} \coreformula{\fretalways}{\cond}{\res}{\phileft}\\
\implies & \rho \models_{n} \historically[{[\lb{I},\ub{I}]}] {\coreformula{\fretalways}{\cond}{\res}{\phileft}}
\end{aligned}$$

By \smartref{cor:baseform_historically}, if $\ub{I} \neq n$ then it holds that
$\rho \models_{\ub{I}+ 1} \baseform{\cond}{\fretalways}{\res}{\phileft}{\phiright}$,
otherwise, it holds that $\rho\models_{\ub{I}} \baseformlast{\cond}{\fretalways}{\res}{\phileft}{\phiright}$.

If $\triggers{\cond}{I}{\rho}\neq\emptyset$, let  $t\in[\lb{I},\ub{I}]$.
If $t<\mathit{min}(\triggers{\cond}{I}{\rho})$, then $\triggers{\cond{}}{\rho}{I} \cap [lb{I},t] = \emptyset$. By \smartref{lem:triggers}, it follows that for all $t_0\in[\lb{I},t]$, $\rho \models_{t_0} \phinotriggers$. From here the proof proceeds as the previous case in which $\triggers{\cond{}}{\rho}{I}=\emptyset$.

Otherwise, if $t\geq \mathit{min}(\triggers{\cond{}}{\rho}{I})$, let $t_0$ be the first trigger $\mathit{min}(\triggers{\cond{}}{\rho}{I})$. By \smartref{lem:triggers}, it follows that $\rho \models_{t_0} \phitrigger$. 
By \smartref{tbl:coreformula}, it follows that for all $j\in[t_0,\ub{I}]$, $j\in\resInt$, and, by \smartref{def:oliexpr}, this means that $\rho \models_j \res$. In particular, $\rho \models_{t_0} \res$, and, thus, 
$\rho \models_{t_0}  \phitrigger \wedge \res$. 
By definition of $\sir{}{}$, it follows that $\rho \models_{t} \sir{\res}{\phitrigger \wedge \res}$ for all $t\in[\lb{I},\ub{I}]$. From here, the result follows from \smartref{cor:baseform_historically} and \smartref{tbl:coreformula}.

Now, consider the $\Leftarrow$ direction.
By \smartref{cor:baseform_historically}, it follows that $\rho \models_{n} \historically[{[\lb{I},\ub{I}]}] \phinotriggers \vee (\sir{\res}{\phitrigger})$.
Let $t\in I$ and split the proof into two cases.

If $\triggers{\cond}{I}{\rho}=\emptyset$ the theorem follows directly.
Otherwise, suppose $\triggers{\cond}{I}{\rho}= \{t_0,t_1,\dots,t_k\}$ such that $t_i< t_{i+1}$ for all $i\in[0,k-1]$. It follows that the interval $[\mathit{min}(\triggers{\cond}{I}{\rho}),\ub{I}]$ can be represented as a union of intervals according to the triggers. That is,
$$[\mathit{min}(\triggers{\cond}{I}{\rho}),\ub{I}] = \bigcup_{i=0}^{k-1} [t_i,t_{i+1}] \cup [t_k,\ub{I}]$$ where $t_0 = \mathit{min}(\triggers{\cond}{I}{\rho})$.

Let $j \in [\mathit{min}(\triggers{\cond}{I}{\rho}),\ub{I}]$ and define $t$ to be either $t_{i+1}$ for $i\in[0,k-1]$ or $\ub{I}$.
By hypothesis, $\rho \models_{t} \phinotriggers \vee (\sir{\res}{\phileft})$, but $\rho \not\models_{t} \phinotriggers$, otherwise by \smartref{lem:triggers}, $\triggers{\cond}{I}{\rho} \cap [\lb{I},t] = \emptyset$, but this is a contradiction since there exists at least one trigger before $t$.
Thus, $\rho \models_{t} \sir{\res}{\phileft}$, and, by definition of $\sir{}{}$, it follows that there exists $r_0$ such that $r_0\leq t$ and $\rho\models_{r_0} \res \wedge \phitrigger$ and for all $r_0\leq r_1 \leq t$, $\rho \models_{r_1}\res$.
By \smartref{lem:triggers}, $\rho \models_{t_i} \phitrigger$ for all $i\in[0,k]$ and, since $j\geq t_i$, it follows that $\rho \models_j \res$, thus, $j\in\resInt$ and the proof is complete.
\end{proof}

The following theorem ensures that the semantics of a \fretish{} requirement is equivalent to the semantics of the past-time MTL formula generated as defined in \smartref{sec:translation}.
That is, if a trace is in the semantics of a \fretish{} requirement, then the generated MTL formula holds at the end of the trace, and viceversa.
\begin{theorem}
  Let $\req$ be a \fretish{} requirement and $\rho\in\TraceDom$, 
  $$\rho \in \fretsem{\req} \iff \rho \models \genform{\req}.$$
\end{theorem}
The proof proceeds by cases on the scope field.
Each proof is split into two main cases to distinguish when the scope is an only scope or not. In the first case, $\fieldtiming$ and $\res$ are replaced with their duals in the proof.
Below, the case in which $\fieldscope = \fretin{\mode}$ is shown.
The \PVS{} formalization contains the lemmas for the other scope fields.
\begin{proof}
Assume that the $\fieldscope$ is $\fretin{\mode}$ for a given $\mode\in\B$.
Let $\rho\in\fretsem{\langle \fretin{\mode},\fieldtiming,\cond,\res \rangle}$ and consider the $\implies$ direction.
First, it is shown that $\rho\models_n \historically[]{(\phibaseform \vee \ftp)}$.
Let $t\in[0,n]$, if $t=0$, then $\rho\models_n \ftp$ and the result follows directly.
If $t\neq 0$ it is possible to distinguish three cases.
If $\modeInt=\emptylin$, it means that there is no scope interval, thus $\rho\not\models_t\phiright$. By Equation~\eqref{eq:baseform}, $\phibaseform$ is of the form $\phiright \ra \phi$ for some $\phi$, thus $\rho\models_t \phibaseform$.
Similarly, if there exists $I\in\modeInt$ such that $t\neq\ub{I}+1$, then, from \smartref{tbl:scopeendpoints} and \smartref{tbl:scopeleftright} $\rho\not\models_t\phiright$ and the proof proceeds in the same way.
Otherwise, if $t=\ub{I}+1$, since $I\in\modeInt$, it follows that $I \in \timeSem{\fieldtiming}{\cond}{\res}$, and, by \smartref{th:baseformeq}, it follows that $\rho\models_{\ub{I}+1} \phibaseform$.
In the following, it is shown that $\rho\models_n (\sir{\neg\phiright}{\phileft}) \ra \phibaseformlast$.
If $\modeInt=\emptylin$, then $\phileft$ is never satisfied and thus, by \smartref{tbl:scopeendpoints} and \smartref{tbl:scopeleftright}, $\rho\not\models_n \sir{\neg\phiright}{\phileft}$, and the theorem follows directly.
If there exists $I\in\modeInt$ such that $\ub{I} = n$, then $\rho\models_n \phibaseformlast$. Otherwise, for all $I$ it holds that $\ub{I}<n$, thus $\rho\models_{\ub{I}} \phiright$ and the result follows directly since $\rho\not\models_n \sir{\neg\phiright}{\phileft}$.

Consider the $\Leftarrow$ direction.
If $\modeInt=\emptylin$, the theorem holds directly.
Otherwise, let $I\in\modeInt$, it is possible to distinguish two cases.
Let $\ub{I}\neq n$. By hypothesis, $\rho\models_n \historically[]{(\phibaseform \vee \ftp)}$, this means that for all $t\in[0,n]$, $\rho\models_{t} \phibaseform \vee \ftp$. Since $\ub{I}+1 \neq 0$, it follows that $\rho\not\models_{\ub{I}+1}\ftp$ and, therefore, $\rho\models_{\ub{I}+1} \phibaseform$. The theorem follows from \smartref{th:baseformeq} since $I \in \timeSem{\fieldtiming}{\cond}{\res}$.
Otherwise, consider $\ub{I}= n$, this means that $I=[\lb{I},n]$.
By hypothesis, it holds that $\rho\models_n (\sir{\neg\phiright}{\phileft}) \ra \phibaseformlast$.
Additionally, $\rho\models_{\lb{I}} \phileft$ and for all $t\in I$, $\rho\models_{\lb{I}} \neg\phiright$, thus, $\rho\models_n \sir{\neg\phiright}{\phileft})$.
In addition, $\rho\models_n \phibaseformlast$, by \smartref{th:baseformeq}, the proof is complete.
\end{proof}

\section{\fretish{} Verification: Lessons Learned}
\label{sec:lesson}
In \cite{GiannakopoulouP21}, a modular and extensible verification framework has been developed to check that the formulas generated by the \fret{} tool conform to the RTGIL semantics of the \fretish{} language.
The framework automatically generates large numbers of example traces using a variety of strategies, in order to cover as many corner cases as possible.
For each generated trace, and for each template combination of \fretish, an oracle produces the expected truth value of requirements corresponding to this combination according to the RTGIL semantics.
This expected value is compared to the value produced when evaluating, with a model checker, the formula computed by \fret{} for this requirement, in the particular trace.
The expected and obtained values are compared, and discrepancies are reported.

As reported in \cite{GiannakopoulouP21}, this framework was extremely valuable in detecting even subtle errors in the translation algorithms during the development of \fret.

Nevertheless, testing, even when extensive, is not proving. And proving, when possible, is extremely valuable for safety-critical applications.
This work developed a robust, compositional proof framework, which significantly
increases the confidence in the current version of the tool and facilitates a continuity of this trust in future extensions of the language.
During the development of this proof framework, the following two oversights were discovered in the algorithm description of \cite{GiannakopoulouP21}.
First, the baseform definition for the timing field $\freteventually$ was defined as $\phicoreformula$.
In the case where the scope field is $\fretafter{\mode}$, the condition is not $\fretnull$, and $\mode$ is never satisfied, the scope is empty, so the requirement should be trivially true.
In examining the actual implementation, it turned out that the definition was instead: $\once[]{\phileft} \ra \phicoreformula$.
Since in this case $\phileft$ is never satisfied, the formula is trivially true, as expected.
The reason this detail was missed in the documentation of the algorithm is that it was hidden in a method invocation that in fact could be simplified.
As a result, that part of the implementation was updated to explicitly match the structure of the algorithm in \cite{GiannakopoulouP21}.
A clear structure that matches the documented algorithms is essential for their maintenance and extensibility.

Additionally, the formula $\ffim{\mode}$ was incorrectly defined as $\fim{\mode} \wedge (\previous{(\historically[]{\neg\mode})})$
instead of $\fim{\mode} \wedge (\ftp \vee \previous{(\historically[]{\neg\mode})})$.
This would exclude the case in which the mode holds immediately at the beginning of the trace ($\ftp$). Again, the implementation was correct and this was an omission in the paper.

The development of a rigorous proof framework in \PVS{} had several additional benefits.
The implementation of the \fretish{} language semantics in \PVS{} was guided by the \fret{}'s diagrammatic explanations and simulation capabilities (see \smartref{fig:gui} and \smartref{fig:OLIs}), and some ambiguities were discovered in the semantics visualization along the way.
In the latest version of \fret{} these ambiguities were resolved, resulting in a clearer user description.

From this research effort, it emerged that the combined use of a rigorous formalization and an extensive testing framework was an excellent strategy for improving the \fret{} tool.
The testing framework was suitable for exploring new ideas, and for providing feedback for discrepancies in an intuitive fashion during the development of \fret{}. 
In addition, it provided an oracle for the \PVS{} semantics implementation.
Even though the \PVS{} formalization was designed to be modular and as compact as possible, the effort of proving all the results is considerable.
So it was essential to start with a semantics that was already tested and did not contain substantial errors.

The rigorous \PVS{} formalization completed the results obtained in the testing phase ensuring the coverage of all corner cases.
Moreover, even though RTGIL diagrams give a more intuitive general picture of the semantics of the language, \PVS{} constructs provide a more explicit enumeration of all special cases that are considered.

The compositional nature of the translation algorithms and proof framework means that it is possible to focus on specific parts of the verification when desired.
For example, it is convenient to start modeling and implementing simple timing operators like $\freteventually$ or $\fretalways$, and gradually expand the implementation to $\fretfor{}$, $\fretafter{}$, $\fretuntil{}$, and $\fretbefore{}$.
Every newly added field value will naturally combine with all the existing ones which are already verified. Any new detected bug would then be easily attributed to the new value.

Finally, the compositional design of the proofs is essential to provide continued theoretical support to \fret{} and to incorporate, with minor effort, a theorem prover based verification step in the \fret{} development process.

\section{Related Work}
\label{sec:related}

The \fretish{} language borrows ideas from the Specification Pattern System (SPS) literature \cite{Patterns}, Easy Approach to Requirements Syntax (EARS, EARS-CTRL) work \cite{EARS,EARSNFM}, and NASA experience.
The SPS work derives a set of patterns from a property specification.
The patterns are structured as a \emph{scope}, which specifies the time intervals where the requirement holds (such as \emph{after} an event or \emph{before} an event), and a \emph{pattern}. The pattern is specified either as an \emph{occurrence} pattern, e.g., specifying that a proposition occurs in the interval, or an \emph{ordering} pattern, e.g., an event being a response to a preceding event.
Each pattern/scope combination is translated into different temporal logics, such as Linear Temporal Logic (LTL) and Graphical Interval Logic (GIL).
This pattern and scope translation to LTL and then to first-order logic has been used for analysis \cite{Redundant2017}.

SpeAR~\cite{FifarekWHR2017} (Specification and Analysis of Requirements) captures requirements that read like natural language.
It supports SPS patterns in its specification language but has reduced these patterns to a very small set, since most of them were never used in practice. Its formal language gets translated to pure past-time LTL.
SpeAR provides a formal logical entailment analysis that proves that specified properties, which define desired behaviors of the system, are consequences of the set of captured assumptions and requirements.
This is done by translating a SpeAR specification into an equivalent Lustre~\cite{JahierRH20} model and analyzing the Lustre model using infinite-state model checking.
This technique provides useful insights on the completeness of the Spear specification.

Other extensions of the SPS include: \emph{real-time property patterns}~\cite{RTSP} which can be translated to metric LTL, Timed CTL, RTGIL, Duration Calculus, Phase Event Automata (PEA) \cite{PostH12}, and to Boogie \cite{Hanfor2019}; \emph{composite propositions}~\cite{Composite,ProspecImproving,Prospec2012}, which address relationships among multiple propositions and are translated to optimized future-time LTL; and \emph{specification of semantic subtleties}~\cite{PROPEL,PROPEL2}.

In \cite{ProspecImproving}, abstract LTL templates are introduced to support automated generation of LTL formulas for complex properties in Prospec~\cite{Prospec2012}. Manual formal proofs and model-checking based testing are used to check that these templates generate the intended LTL formulas.
In contrast with the work presented in this paper, the proofs in \cite{ProspecImproving} are done by hand and are not formalized in a theorem prover. In addition, the structure of the proposed LTL templates is not compositional, therefore it is necessary to prove every single pattern combination.

FORM-L~\cite{FORM-L} is a formal requirements language implemented with Extended Temporal Logic (ETL)~\cite{ETL}. The latter has relevant constructs (e.g., constraints on the number of events occurring), but is manually translated into requirement monitors expressed using blocks in a Modelica library.

Several commercial tools are also available for formal requirements engineering.
The ASSERT tool~\cite{ASSERT,ASSERT2}, proprietary to GE, uses an ontology-based approach both for formalizing domains through the language SADL, and the requirements themselves, through the language SRL. Requirements are in the form of assignments to attributes conditioned on (possibly temporal) Boolean conditions. 
The STIMULUS tool~\cite{STIMULUS} enables its user to specify a formal requirement by combining phrases from a library whose underlying semantics are hierarchical state machines and dataflow constraints.
The phrases can specify metric temporal conditions. The behavior of the resulting set of requirements can be simulated allowing the user to observe the system behavior as specified.
The BTC Embedded\textit{Platform}\textsuperscript{\textregistered} tool~\cite{BTC} provides a GUI to construct requirements according to a graphical ``simplified universal pattern". The pattern consists of a trigger and an action, both with specified events, conditions, and timing constraints. The requirements can then be analyzed, using model-checking, for consistency, completeness, and correctness, and tests can be automatically generated.

Examples of tools that parse and formalize more general natural language are VARED~\cite{VARED} and ARSENAL~\cite{ARSENAL}.
VARED aimed at translating natural language to LTL via SALT~\cite{BauerL11}.
SALT translates to LTL via rewriting rules, but, to the best of the authors' knowledge, it doesn’t prove that the transformation preserves any kind of semantics.
ARSENAL translates natural language to SAL~\cite{SAL} models.

To the best of the authors' knowledge, \fret{} is the only structured natural language elicitation tool that is supported by a rigorous formalization in a theorem prover.

\section{Conclusion}
\label{sec:conclusion}

This paper presents a formalization of the \fretish{} structured natural language and a proof of the correctness of the past-time MTL formula generation algorithm implemented in \fret{}.
This proof improves the confidence in the \fret{} tool for its use in the requirement elicitation of safety-critical systems.

The proposed formalization provides a rigorous theoretical basis to support current and future \fretish{} features.
The modular structure of the proof has been designed to facilitate the
task of extending the language while also maintaining the correctness of the formula generation algorithm.
For example, just one new lemma is needed for any new timing or scope field construct.

In \cite{GiannakopoulouP21}, the equivalence between \fret{} generated future-time and past-time formulas is checked on finite traces of specified length.
As future work, the authors plan to prove the correctness of the algorithm also for future-time formulas over traces of \emph{arbitrary} length.
This proof can be done by leveraging the semantic equivalence between \fretish{} and past-time \MTL{} shown in this paper.
In this way, it will be sufficient to show that the future-time and past-time temporal formulas generated for a given requirement are semantically equivalent.

Support for infinite trace semantics has been recently added to \fret{} in \cite{GiannakopoulouP21}.
The \PVS{} formalization can be extended in the future to target infinite traces by replacing bounded \fretish{} traces with unbounded ones.
Finally, the proposed formalization will be extended as new constructs and features are added to the \fretish{} language, keeping, in this way, a robust theoretical framework behind \fret.

\bibliography{biblio}

\end{document}